\documentclass[lettersze]{IEEEtran}
\usepackage{amsmath,amsfonts}

\usepackage{array}
\usepackage{textcomp}
\usepackage{stfloats}
\usepackage{url}
\usepackage{verbatim}
\usepackage{graphicx}
\usepackage{cite}
\usepackage{amssymb}
\usepackage{subfigure}
\usepackage{caption}
\usepackage{array}
\usepackage{epstopdf}
\usepackage{subfigure}
\usepackage[strings]{underscore}
\usepackage{comment}

\usepackage[ruled,vlined]{algorithm2e}

\newtheorem{definition}{Definition}
\newtheorem{proposition}{Proposition}
\newtheorem{assumption}{Assumption}
\newtheorem{theorem}{Theorem}

\begin{document}

\title{Deep Learning-Based Channel Extrapolation for Pattern Reconfigurable Massive MIMO}

\author{Mu Liang, Ang Li

	}

\maketitle

\begin{abstract}
Reconfigurable antennas that can dynamically change their operation state exhibit excellent adaptivity and flexibility over traditional antennas, and MIMO arrays that consist of multifunctional and reconfigurable antennas (MRAs) are foreseen as one promising solution towards future Holographic MIMO. Specifically, in pattern reconfigurable MIMO (PR-MIMO) communication systems, accurate acquisition of channel state information (CSI) of all the radiation modes is a challenging task, because using conventional pilot-based channel estimation techniques in PR-MIMO systems incurs overwhelming pilot overheads. In this letter, we leverage deep learning methods to design a PR neural network, which can use the estimated CSI for one radiation mode to infer CSIs for the other radiation modes. In order to reduce the pilot overheads, we propose a new channel estimation method specially for PR-MIMO systems, which divides the transmit antennas of PR-MIMO into groups and antennas in different groups employ different radiation modes. Compared with conventional full-connected real-valued deep neural networks (DNN), the PR neural network which uses complex-valued coefficients can work directly in the complex domain. Experiment results show that the proposed channel extrapolation method offers significant performance gains in terms of extrapolation accuracy over benchmark schemes.  
\end{abstract}


\section{Introduction}
\IEEEPARstart{M}{ultifunctional} and reconfigurable antennas (MRAs) are capable of adapting their working state such as frequency, polarizations and radiation patterns compared with conventional arrays, and are therefore foreseen as promising candidates for 6G\cite{2015Fundamentals}. It has been shown in \cite{2008Design} that using MRAs can increase the spectral efficiency of MIMO communication systems, because MRAs bring additional degrees of freedom and can change the current distribution of the antenna array and the resulting electromagnetic field. As for the implementation of MRAs, existing solutions include semi-conductor switches, liquid metals, micro-electro-mechanical systems (MEMS) switching , metasurface, etc\cite{2003Microstrip}.\par

The performance of MIMO systems is highly dependent on the link quality between the transmitter and the receiver, which are determined by the angles of departure, angles of arrival, path gains and the array related coefficients such as antenna gain and directivity. These parameters are determined by the physical characteristics of wireless environments and the radiation mode of the antenna array. Compared with conventional array with fixed radiation mode, communication systems using MRAs have the potential to achieve improved performance by dynamically altering the radiation mode. In \cite{2018Low}, it is shown that the performance of MIMO systems with MRAs will improve as the number of reconfigurable modes increases, while the performance gains become less significant when the number of reconfigurable modes is large.\par 
One important problem existing in pattern reconfigurable MIMO (PR-MIMO) communication systems is how to choose a promising radiation mode during data transmission. In the literature, most solutions to date are dependent on the multi-armed bandit or Thompson sampling theory \cite{2011Analysis,2014Learning,2021Online}. For example, in\cite{2014Learning}, the PR mode selection problem is formulated as a multi-armed bandit problem to select the best radiation pattern via an online learning process without the need for instantaneous channel state of information (CSI) of all radiation modes. In \cite{2021Online}, a Thompson sampling algorithm with channel prediction (TS-CP) is proposed based on the multi-armed bandit theory. The TS-CP exploits channel correlation to predict the channel conditions of unexplored radiation modes.\par

The above mode selection approaches require full/partial CSI information of the PR-MIMO systems, while the CSI acquisition for PR-MIMO systems is a challenging task, this is because alterations on the radiation mode of a single antenna will change the radiation mode of the array, and excessive pilot overheads will be required if the transmitter aims to acquire the channel information for all radiation modes via traditional channel estimation approaches. Existing solutions to the above problem includes \cite{5585633} which only selects a subset of reconfigurable modes with traditional pilot-based channel estimation, while ignoring the other reconfigurable modes, resulting in decreased reconfigurability gain. 
In addition, Bahceci et al \cite{2017Efficient} proposes a Gram–Schmidt based decomposition algorithm that utilizes the correlation between different reconfigurable modes to predict the CSI to reduce the pilot overheads. The algorithm proposed in \cite{2017Efficient} requires perfect knowledge of channel statistics, which is unattainable and results in pilot overheads that are at least twice the amount for conventional MIMO systems. As reviewed, the above approaches cannot fully exploit the performance benefits of PR-MIMO systems or return satisfactory extrapolation performance. On the other hand, recently deep learning based approaches has been extensively studied for MIMO channel estimation to reduce signaling overheads \cite{2018Deep,2019Deep,2020Deep,2019Deepfnn}. In \cite{2018Deep}, the authors use the convolutional neural network to compress the CSI feedback, which can greatly reduce the amount of feedback information needed. In \cite{2019Deep}, a sparse neural network (SCNet) that uses the uplink CSIs to extrapolate the downlink CSIs in FDD massive MIMO syetems is proposed. In \cite{2020Deep}, the authors use a convolutional neural network to extrapolate the CSIs in TDD MIMO systems. In \cite{2019Deepfnn}, the full-connected deep neural network (DNN) proposed for the channel calibration in massive MIMO systems, achieving better channel estimation performance than conventional methods. To the best of our knowledge, deep learning based approaches have not been fully exploited for PR-MIMO.\par

In this paper, we propose a new channel estimation method that divide the antenna elements into groups where antennas in different groups employ different radiation modes during channel estimation, and leverage deep learning based solutions to address the signaling overhead issue during the channel estimation process for PR-MIMO systems. The proposed channel estimation method enjoys the same signaling overhead as for conventional MIMO systems. We not only exploit the deterministic estimated-to-extrapolated channel mapping function for PR-MIMO systems and prove its existence, but also demonstrate how the neural network can be trained to fit the channel mapping function. Building upon this, we propose a PR-Net that can extrapolate the CSI of all radiation modes without introducing any additional overheads. Numerical results demonstrate that the complex-valued PR-Net can outperform the real-valued DNN in channel extrapolation task with low pilot overheads, both outperforming the existing channel extrapolation method for PR-MIMO systems.\par

\emph{Notations:}
$\alpha$, $\boldsymbol{\alpha}$, $\boldsymbol{A}$ denote scalar, vector and matrix, respectively. $vec\left ( \cdot  \right )$ denotes the vector form of a matrix, $\Re[\cdot]$ and $\Im[\cdot]$ extract the real and imaginary parts of the argument, respectively.
$\left \| \boldsymbol{\cdot} \right \| _{2} $ denotes the Euclidean norm. $(\cdot)^H$  denotes the conjugate transposition. $ \circ $ represents the composite mapping operation. $ \mathbb{E}\left [ \cdot  \right ] $ represents the expectation. $\rightarrow $ represents the mapping operation.

\section{System Model} 
We consider a point-to-point PR-MIMO (P2P-PR-MIMO) system, which the transmitter is equipped with a MRA of $M\gg1$ reconfigurable antennas and the receiver is equipped with $N>1$ conventional omnidirectional antennas. We assume that the number of reconfigurable modes is $P$, where $P\ll M $ for massive MIMO scenarios. Each antenna of the MRAs can change its reconfigurable mode independently, and for simplicity uniform linear arrays are employed at both sides. We consider a multi-path channel model as in \cite{Zhihong2003Spatial}, where channel $\boldsymbol{H}^{p}\in\mathbb{C}^{N\times M}$ for the $p$-th reconfigurable mode between the receiver and the transmitter can be expressed as:

\begin{equation}
	\boldsymbol{H}^{p} = \frac{1}{\sqrt{MN}} \sum_{i=1}^{N_{cl}}\sum_{j=1}^{N_{ray}}\alpha _{i,j}\cdot\beta _{i,j}(\theta_{i,j},\phi _{i,j} )
	\boldsymbol{\alpha}_{r}(\theta_{i,j})\boldsymbol{\alpha} _{t}^{H}(\theta_{i,j}),  
\end{equation}

\noindent
where $N_{cl}$, $N_{ray}$, $\alpha_{i,j}$, $\theta_{i,j}$ and $\phi_{i,j}$ are the number of clusters, the number of rays in each cluster, complex path gain, azimuth angle of departure of the $i$-th cluster and the $j$-th ray, elevation angle of departure of the $i$-th cluster and the $j$-th ray, respectively. $\beta_{i,j}^{p}(\theta_{i,j},\phi_{i,j})$ is the reconfigurable mode gain which is determined by the $\theta_{i,j}$ and $\phi_{i,j}$. Moreover, $\boldsymbol{\alpha}(\theta_{i,j})$ is the array steer vector defined as:
\begin{equation}
	\boldsymbol{\alpha} (\theta_{i,j})  = [1,e^{-j\chi\sin \theta_{i,j}},\cdot\cdot \cdot , e^{-j\chi(M-1)\sin \theta_{i,j}}]^{T}, 
\end{equation}
\noindent
where $\chi = 2\pi df/c$, $d$ is antenna spacing, $c$ is the speed of the light, $f$ is carrier frequency. $\theta$ and $\phi$ follow a uniform discrete distribution in $[0,2\pi]$ and $[0,\pi] $, respectively.

\begin{figure}[!t]
	\centering
	{\includegraphics[width=1.02\linewidth]{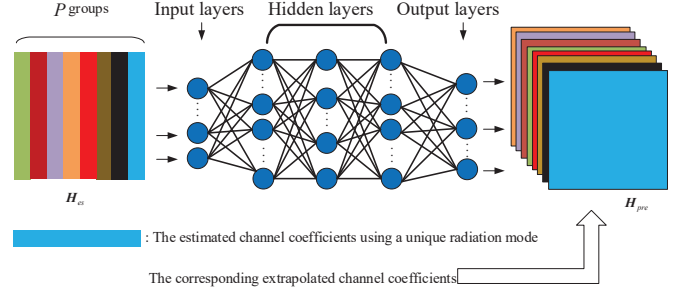}}
	\captionsetup{labelformat=default,labelsep=space}
	\caption{An illustration for the proposed channel estimation and extrapolation method for PR-MIMO}
	\label{fig.1}
\end{figure}

\section{Channel Estimation and Mapping Formulation}

In this section, we introduce the proposed channel acquisition methods for PR-MIMO systems. To begin with, we firstly introduce a novel channel estimation scheme for PR-MIMO systems with low pilot overheads, based on which we formulate the channel mapping function following \cite{2020Deep}. Finally, we use deep learning to find the channel mapping function for channel extrapolation.

\subsection{Channel Estimation for PR-MIMO}

First, we divide all the antennas at the transmitter into $P$ groups with each group consisting of $\left \lfloor M/P \right \rfloor $ antennas, and antennas in the same group employ the same radiation mode during channel estimation, as shown in Fig. 1. 
In particular, if $mod(M,P) \ne 0$, the last group contains $mod(M,P)$ antenna elements.
Subsequently, the transmitter sends a sequence of pilot signals to estimate the channel, and in this way the pilot overheads remain the same as the conventional MIMO systems. The received signal can be expressed as:
\begin{equation}
	\boldsymbol{Y} = \boldsymbol{H}\boldsymbol{X} + \boldsymbol{W},
\end{equation}
where $\boldsymbol{Y}\in \mathbb{C}^{N \times M} $  is the received signal matrix, we define the length of the pilot sequence to be equal to the number of transmitter antennas, $\boldsymbol{X} \in \mathbb{C}^{M \times M}$ is the pilot matrix that satisfies $\mathbb{E}\left [ \boldsymbol{X}\boldsymbol{X}^{H} \right ] = \boldsymbol{I}_{M}$, and $\boldsymbol{W}\in \mathbb{C}^{N \times M}$ represents noise matrix whose elements follow complex Gaussian distribution with zero mean and variance $\sigma^{2}$. We define $\boldsymbol{H}_{es}\in \mathbb{C}^{N \times M}$ as the estimated channel matrix, and by using linear-minimum-mean-squared-error (LMMSE) estimator, $\boldsymbol{H}_{es}$ can be obtained as:
\begin{equation}
	\boldsymbol{H}_{es} = \boldsymbol{Y}(\boldsymbol{X}^{H}\boldsymbol{R}_{H}\boldsymbol{X}+\sigma ^{2}N\boldsymbol{I}_{M})^{-1}\boldsymbol{X}^{H}\boldsymbol{R}_{H}, 
\end{equation}
where $\boldsymbol{R}_{H}=\mathbb{E}\left[ \boldsymbol{H}^{H}\boldsymbol{H} \right ]\in\mathbb{C}^{M\times M}$. For ease of subsequent derivations, we further introduce $\boldsymbol{h}_{es} = vec(\boldsymbol{H}_{es})\in\mathbb{C}^{MN\times 1}$ as the vector form of $\boldsymbol{H}_{es}$, $\boldsymbol{H}_{all}\in \mathbb{C}^{N \times M \times P}$ as the collected CSI for all reconfigurable modes, $\boldsymbol{h}_{all} = vec(\boldsymbol{H}_{all})\in\mathbb{C}^{M*N*P\times 1}$ as the vector form of $\boldsymbol{H}_{all}$, $\boldsymbol{H}_{pre} \in \mathbb{C}^{N \times M \times(P-1)}$ as the CSI that needs to be extrapolated, and $\boldsymbol{h}_{pre}=vec(\boldsymbol{H}_{pre})\in \mathbb{C}^{N*M*(P-1)\times 1}$ as the vector form of $\boldsymbol{H}_{pre}$.

\subsection{Existence of Channel Mapping Function for PR-MIMO}

Compared with traditional MIMO systems where the CSI is solely dependent on the wireless propagation environment, the CSI for PR-MIMO systems is dependent both on the wireless propagation environment and the radiation modes at the transmitter/receiver. Therefore, given the location of the receiver and the transmitter, when the radiation modes of the antenna array remain fixed, there should exist a deterministic mapping function between the receiver and the transmitter within a channel coherence interval. For simplicity of description, we define the $\boldsymbol{h}(p)$ as the vector form of $\boldsymbol{H}^{p}$.\par

\begin{definition}\label{d1}
The mapping function $\boldsymbol{\Psi}$ of the $p$-th radiation mode between channel vector $\boldsymbol{h}(p)$ and the receiver location can be described as follow:
\begin{equation}
	\boldsymbol{\Psi}:{(D,\theta,\phi,f )}\to {\boldsymbol{h}(p)},
\end{equation}
\end{definition}

\noindent
where $D$ is the distance between transmitter and receiver, ${(D,\theta,\phi,f )}$ and $\boldsymbol{h}(p)$ represent the domain and codomain of the mapping function.\par

Antennas at the transmitter can be divided into two sets based on whether the channel coefficients are estimated or extrapolated for the $p$-th radiation mode. We define these two sets as the estimated channel coefficients set $\mathcal{M}_{1}$ and the extrapolated channel coefficients set $\mathcal{M}_{2}$, respectively. 
Let $h^{nm,p}_{es}\in \mathcal{M}_{1}$ represent the estimated channel coefficient of the $p$-th radiation mode from the $n$-th receiver antenna to the $m$-th transmitter antenna, and let $h^{nl,p}_{pre}\in \mathcal{M}_2$ represent the extrapolated channel coefficient of the $p$-th radiation mode from $n$-th receiver antenna to the $l$-th transmitter antenna, respectively. 
We can naturally draw a conclusion that there exists a mapping relationship between the two sets, because they experience the same wireless propagation environment, based on which we define the reconfigurable antenna mapping function $\boldsymbol{\Psi}_{\mathcal{M}}$.\par 
\begin{definition}\label{d2}
The reconfigurable antenna mapping function $\boldsymbol{\Psi} _{\mathcal{M}}$ between $\mathcal{M}_{1} $ and $\mathcal{M}_{2} $ can be described as:
\begin{equation}
	\boldsymbol{\Psi}_{\mathcal{M}}:{\boldsymbol{h}_{es}(\mathcal{M}_{1},p)}\to {\boldsymbol{h}_{pre}(\mathcal{M}_{2},p)},    p=1,2,...,P,
\end{equation}
\end{definition}

\noindent
Next, we adopt the following assumption.\par

\begin{assumption}
	\label{a1}
The position-to-channel mapping function $\boldsymbol{\Psi}:{(D,\theta,\phi,f )}\to {\boldsymbol{h}(p)}$ and estimated-to-extrapolated channel mapping function $		\boldsymbol{\Psi}_{\mathcal{M}}:{\boldsymbol{h}_{es}(\mathcal{M}_{1},p)}\to {\boldsymbol{h}_{pre}(\mathcal{M}_{2},p)}, p=1,2,...,P$ are bijective.
\end{assumption}

\emph{Assumption} \emph{\MakeUppercase{\romannumeral 1}} means that channel coefficients $h^{nm,p}_{es}$ and $h^{nl,p}_{pre}$ have a unique relationship with the location of the transmitter and receiver. Although this cannot be proved analytically, the probability that $\boldsymbol{\Psi}$ and $\boldsymbol{\Psi}_{\mathcal{M}}$ are bijective is close to 1 in practical wireless communication
scenarios\cite{2019Deep}.\par
By combining the position-to-channel mapping function and the estimated-to-extrapolated channel mapping function, we arrive at the PR-MIMO channel mapping function, as presented in \emph{Proposition} \emph{\MakeUppercase{\romannumeral 1}} below:\par

\begin{proposition}
	\label{p1}
	 According to \emph{Assumption} \emph{\MakeUppercase{\romannumeral 1}}, the PR-MIMO channel mapping function can be written as:
	 \begin{small}

	\begin{equation}
		\boldsymbol{\Phi} = \boldsymbol{\Psi}\circ \boldsymbol{\Psi}_{\mathcal{M}}:{\boldsymbol{h}_{es}(D,\theta,\phi,f,\mathcal{M}_{1},p)}\to {\boldsymbol{h}_{pre}(D,\theta,\phi,f,\mathcal{M}_{2},p)}, 
	\end{equation}
     \end{small}
\end{proposition}
where $\boldsymbol{\Psi}\circ \boldsymbol{\Psi}_{\mathcal{M}}$ represents the composite function.\par

\emph{Proof :} From the \emph{Definition} \emph{\MakeUppercase{\romannumeral 1}} and \emph{Definition} \emph{\MakeUppercase{\romannumeral 2}}, we have the function $\boldsymbol{\Psi}:{(D,\theta,\phi,f)}\to {\boldsymbol{h}(p)}$ and $\boldsymbol{\Psi}_{\mathcal{M}}:{\boldsymbol{h}_{es}(\mathcal{M}_{1},p)}\to {\boldsymbol{h}_{pre}(\mathcal{M}_{2},p)}$. At the same time, they are all bijective under \emph{Assumption} \emph{\MakeUppercase{\romannumeral 1}}. Therefore, the composite mapping $\boldsymbol{\Psi}_{f}\circ \boldsymbol{\Psi}_{\mathcal{M}}$ exists for any possible channel parameters $(D,\theta,\phi,f,p)$ and channel coefficients set $\mathcal{M}_{1}$, $\mathcal{M}_{2}$.

\subsection{Deep Learning for Channel Mapping}
We have introduced the channel mapping function for PR-MIMO systems in \emph{Proposition} \emph{\MakeUppercase{\romannumeral 1}}. However, this relationship is difficult to describe using the traditional mathematical formulation, which motivates us to use deep learning based approach to construct such relationship. According to \cite{1989Multilayer}, \emph{Theorem} \emph{\MakeUppercase{\romannumeral 1}} is presented below:

\begin{theorem}
For any given small error $\varepsilon > 0$, there always
exists a positive constant $K$ that is large enough such that
\begin{equation}
	\sup _{\boldsymbol{x}\in \boldsymbol{h}_{all}}\left \| NET_{K}(\boldsymbol{x},\boldsymbol{\Omega})-\boldsymbol{\Phi} \right \|\le \varepsilon ,\boldsymbol{h}_{all}={\boldsymbol{h}(D,\theta,\phi,f,\mathcal{M})}, 
\end{equation}	
\end{theorem}
where ${NET}_{K}(\boldsymbol{x}, \boldsymbol{\Omega})$ is the output of a multiple-layer feedforward network with the input data $\boldsymbol{x}$, where $\boldsymbol{\Omega}$ and $K$ represent the network parameters and the layer numbers.\par 

To summarize, acquisition of the full CSI for PR-MIMO systems consists of 3 steps: 1) divide antennas into groups with different group employing different radiation modes, 2) use traditional channel estimation method to obtain $\boldsymbol{H}_{es}$, 3) use a complex-valued PR-Net to extrapolate full CSI $\boldsymbol{H}_{all}$, as presented in Algorithm 1. In Section IV below, we describe the complex-valued PR-Net in detail.\par

\begin{algorithm}[ht]
	\SetKwInOut{KIN}{input}
	\SetKwInOut{KOUT}{output}
	\caption{Channel Estimation and Extrapolation for PR-MIMO Systems}
	\KIN{pilot matrix $\boldsymbol{X}$, learning rate $\gamma$, batch size $R$, radiation number $P$, epochs number $E$}
	\KOUT{full CSI $\boldsymbol{H}_{all}$}
	Randomly initialize the complex network parameters $\boldsymbol{\Omega}$;\par
	\For{$t=1,\dots, E$}{Update the $\boldsymbol{\Omega}$ by using the ADAM algorithm (learning rate $\gamma$) to minimize the $\mathrm{Loss}$;}\par
	Divided the transmitter antenna into $P$ groups;\par 
	Allocate radiation pattern to each antenna group;\par
	Send the $\boldsymbol{X}$ and get $\boldsymbol{H}_{es}$ using (4);\par
	Reshape the $\boldsymbol{H}_{es}$ as $\boldsymbol{h}_{es}$;\par
	Obtain $\boldsymbol{h}_{pre}$ based on (9);\par
	Reshape $\boldsymbol{h}_{pre}$ and $\boldsymbol{h}_{es}$ as $\boldsymbol{H}_{all}$.
\end{algorithm}

\section{PR-Net for Channel Extrapolation}
In this section, we first introduce the complex-valued PR-Net, followed by the discussion on its training and deployment in PR-MIMO systems to acquire all the CSIs based on the proposed channel estimation approach.
\subsection{PR-Net Architecture}

As shown in Fig. 1, the PR-Net operates in the complex-valued domain, which means that both the input $\boldsymbol{h}_{es}$ and the output $\boldsymbol{h}_{pre}$ can be directly incorporated into the network. Specifically, $\boldsymbol{h}_{pre}$ contains all channel coefficients of all radiation modes. The number of neurons in each hidden layer remains same. A cascade of nonlinear transformation of $\boldsymbol{h}_{es}$ can be expressed as:
\begin{equation}
\boldsymbol{h}_{pre}= \boldsymbol{f}^{(K-1)}(\cdots \boldsymbol{f}^{(1)}(\boldsymbol{h}_{es}{(D,\theta ,\mathcal{M}_{1},f,p )})),
\end{equation}	
where $\boldsymbol{f}$ is the nonlinear transformation function of the layer and can be described as:

\begin{equation}
	\boldsymbol{f}^{(k)}(\boldsymbol{x}) =  \left\{\begin{array}{ll}
		\boldsymbol{g}\left(\boldsymbol{W}^{(k)} \boldsymbol{x}+\boldsymbol{b}^{(k)}\right),  1 \leq k<K-1 \\
		\boldsymbol{W}^{(k)} \boldsymbol{x}+\boldsymbol{b}^{(k)},  k =  K-1
	\end{array}\right.
\end{equation}
\noindent
where $\boldsymbol{g}$ is the activation function and is given as:
\begin{equation}	
	\boldsymbol{g}(\boldsymbol{z}) = max\left \{ \Re \left [ \boldsymbol{z} \right ]  ,\boldsymbol{0}\right \} +j\cdot max\left \{ \Im  \left [ \boldsymbol{z} \right ]  ,\boldsymbol{0}\right \},
\end{equation}
with the notations $\Re$[$\cdot$] and $\Im$[$\cdot$] being the real and imaginary parts of the vectors, respectively.\par

In \cite{1989Multilayer}, it is proven that a three-layer
network can approximate an arbitrary function, while our proposed PR-Net has more than three layers in order to increase accuracy. Compared with the real-valued DNN networks, the complex-valued PR-Net has higher adaptability and is more suitable in learning complex-valued functions. At the same time, the proposed PR-Net reduces the number of neurons in the hidden layers to decrease the complexity. As a result, the PR-Net can not only reduce the amounts of network parameters but also offer higher extrapolation accuracy.\par

\begin{figure*}[b]
	\centering
	\begin{minipage}[t]{0.45\textwidth}
		\centering
		\includegraphics[width=\textwidth]{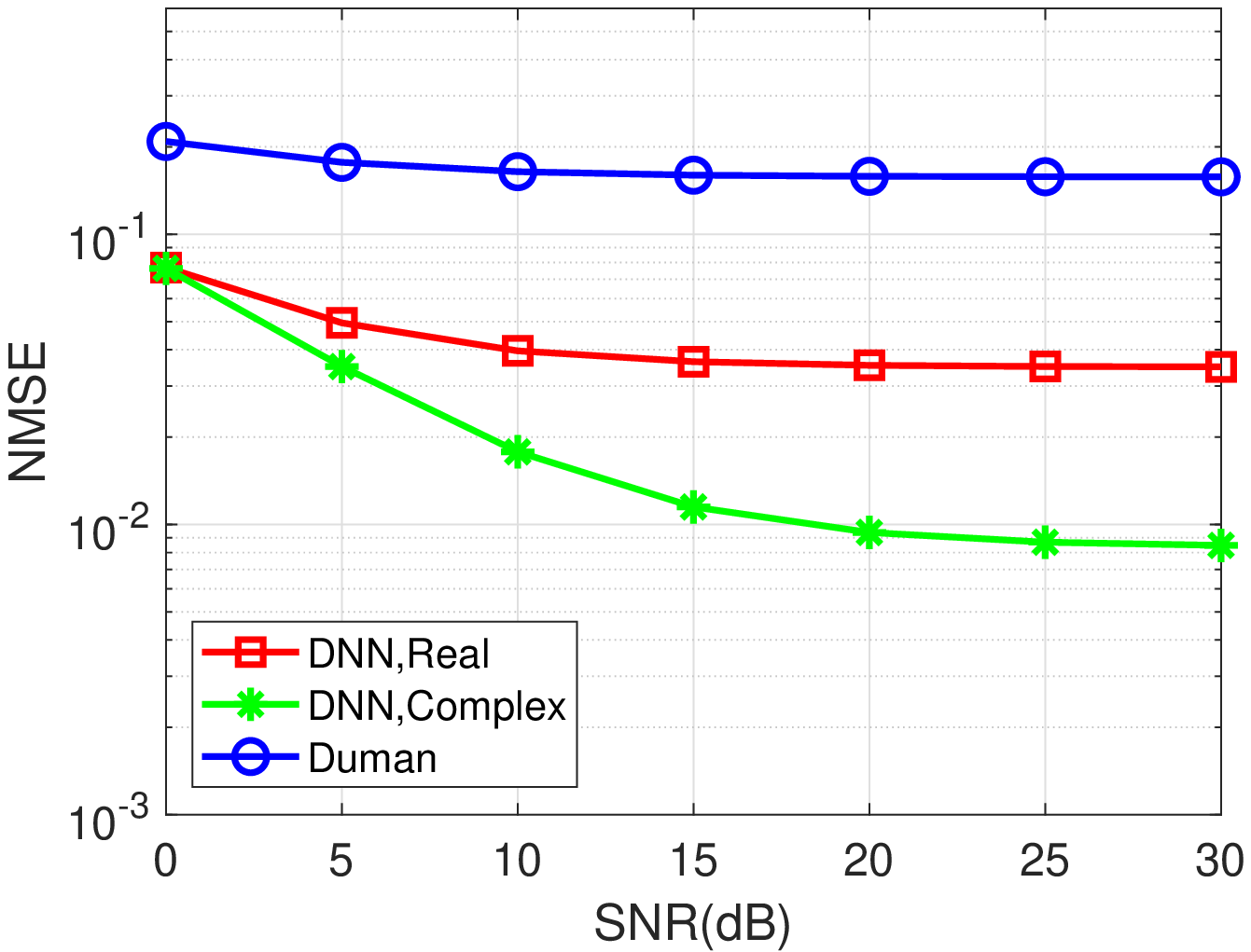}
		\captionsetup{labelformat=default,labelsep=space}
		\caption{NMSE v.s. SNR, $M=64$, $N=8$, $P=8$}
		\label{fig.2}
	\end{minipage}\hfill
	\begin{minipage}[t]{0.45\textwidth}
		\centering
		\includegraphics[width=\textwidth]{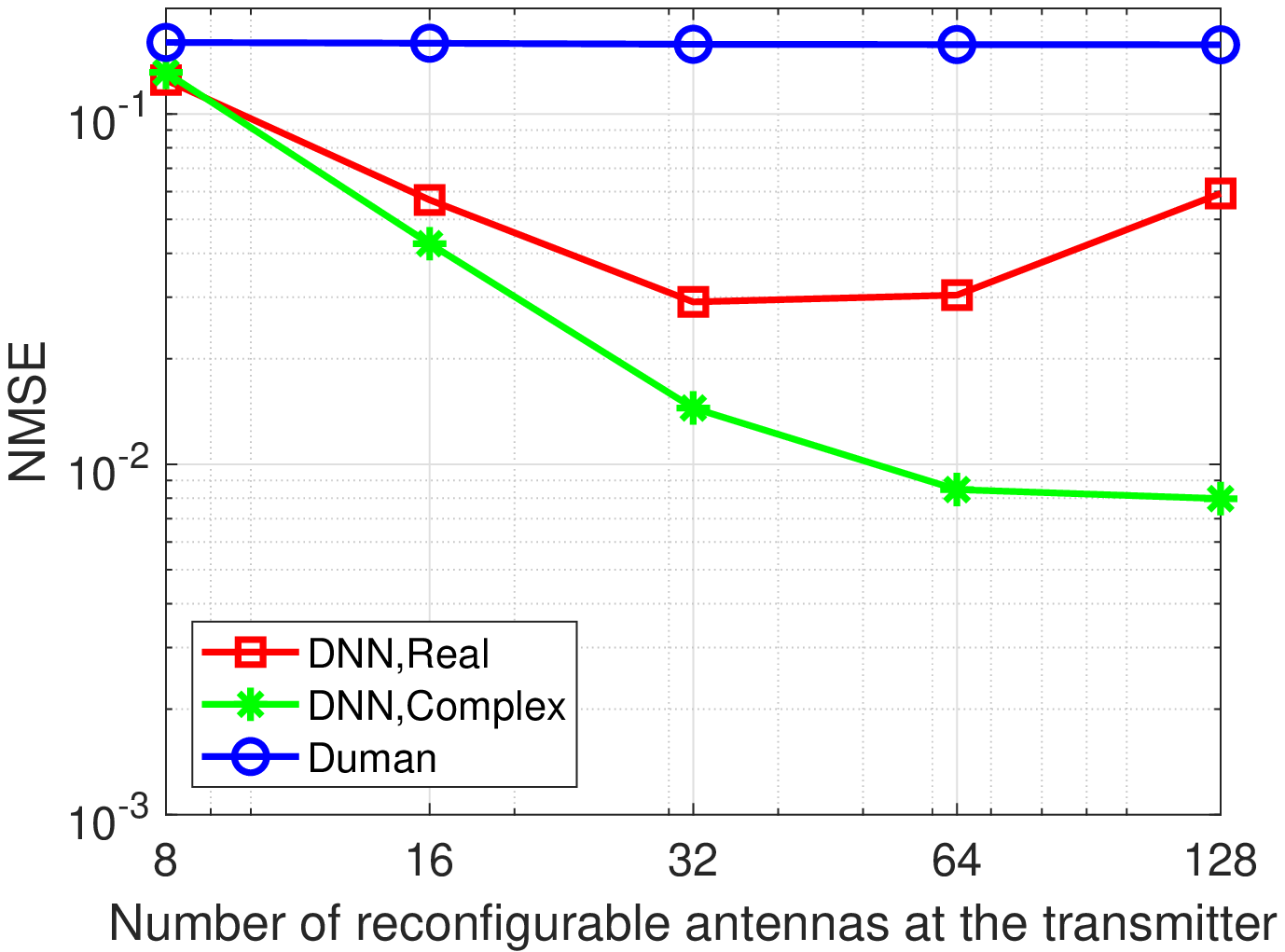}
		\captionsetup{labelformat=default,labelsep=space}
		\caption{NMSE v.s. $M$, SNR=30 dB, $N=8$, $P=8$}
		\label{fig.3}
	\end{minipage}
\end{figure*}

\subsection{Training and Depolyment}

There are two stages to deploy PR-Net to extrapolate the reconfigurable modes channel: the off-line training and the on-line extrapolation. In the off-line training stage, we divide the transmitter antennas into $P$ groups, which correspond to the radiation modes, and employ the conventional channel estimation method to obtain the estimated channel matrix. The transmitter collects these estimated channel matrices as training samples to train the PR-Net. In the extrapolation stage, the parameters of the PR-Net are fixed. We reshape the output of the PR-Net to obtain the channel matrices for all radiation modes. we define the noisy output of the network as $\hat{\boldsymbol{h}}_{pre}$. The PR-Net is trained to minimize the difference between $\hat{\boldsymbol{h}}_{pre} $ and the noiseless supervise label $\boldsymbol{h}_{pre}$. The loss function is defined as:

\begin{equation}	
\mathrm{Loss}=\frac{1}{RL_{h}}\sum_{r=0}^{R-1}\left \| \hat{\boldsymbol{h}}_{pre}-\boldsymbol{h}_{pre} \right \|_{2}^{2},  
\end{equation}

\noindent
where $R$ is the batch size, the superscript ($r$) denotes the index of the $r$-th training sample, and $L_h$ is the length of the vector $\boldsymbol{h}_{pre}$. The loss function is minimized by the adaptive moment estimation (ADAM) algorithm until the PR-Net converges.\par  

\section{Simulation Results}
In this section, we present numerical validations of the proposed channel estimation and extrapolation method for PR-MIMO. Unless otherwise specified, the system parameters are set
as follows: the transmitter is equipped with $M=64$ MRAs; each antenna has $P=8$ reconfigurable antenna modes; the receiver has $N=8$ conventional omni antennas; the working frequency $f = 2.5$ GHz; the number of clusters and rays is $N_{cl}=10$ and $N_{ray}=20$, respectively; each path gain $\alpha_{i,j}$ follows a standard complex Gaussian distribution.\par

To demonstrate the effectiveness of the deep learning neural network, we compare the channel extrapolation accuracy with the algorithm proposed in \cite{2017Efficient} (denoted as 'Duman'), which requires sending the pilot matrix multiple times with increased overheads. We also use the real-valued DNN as a benchmark, which was originally designed for uplink/downlink channel calibration for MIMO systems. Throughout the simulations, we employed Tensorflow 2.1 as the deep learning framework for the PR-Net, with the number of neurons in the hidden layers set to (512, 512, 512). The initial learning rate of the ADAM algorithm is 0.001, and the batch size is 32. The parameters of the $k$-th layer of PR-Net are initialized as a complex Gaussian distribution with zero mean and variance $1/n_{k}$, where $n_{k}$ represents the number of neurons in the $k$-th layer. During the training stage, we used the LMMSE algorithm to estimate the CSIs fed to the PR-Net when the signal-to-noise ratio (SNR) is 25 dB. The number of training samples is 10240, and we conduct 500 epochs. The simulation is carried out with a test size of 0.4, and the channel extrapolation accuracy was measured using normalized-mean-squared-error (NMSE), which is defined as:

\begin{equation}	
	\mathrm{NMSE} = \mathbb{E}\left [ \left \| \boldsymbol{h}_{Pre}-\hat{\boldsymbol{h}}_{pre}  \right \|_{2}^{2} /  \left \| \boldsymbol{h}_{pre} \right \|_{2}^{2}    \right ].
\end{equation}

\begin{figure}[!t]
	\centering
	{\includegraphics[width=0.45\textwidth]{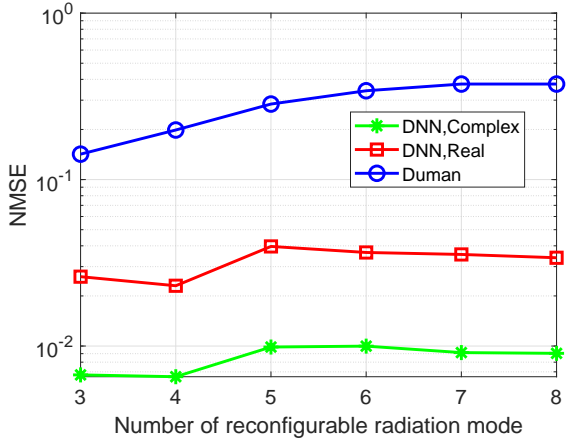}}
	\captionsetup{labelformat=default,labelsep=space}
	\caption{NMSE v.s. $P$, $M=64$, SNR=30 dB, $N=8$}
	\label{fig.4}
\end{figure}

Fig. 2 presents the average NMSE performance of the PR-Net, DNN, and Duman algorithm. The results indicate that the deep learning methods perform better than the Duman algorithm in terms of channel extrapolation accuracy. Moreover, the complex-valued PR-Net has superior performance to the real-valued DNN, demonstrating the capability of complex-valued PR-Net to better fit complex functions. As the SNR increases, the NMSE curve decreases, and in the high SNR region, the slope of the curve gradually becomes flat.

Fig. 3 illustrates the relationship between the NMSE performance and the number of MRAs at the transmitter, given a fixed SNR of 30 dB. The results demonstrate that the accuracy of channel extrapolation using the complex-valued PR-Net improves as the number of antennas increases, while the performance of the real-valued DNN degrades when the number of antennas becomes too large. One possible reason for this phenomenon is that with a larger number of antennas, more information about radiation modes can be utilized to extrapolated the channel matrix. However, as the number of antennas increases, the dimensions of the output also increase. Due to the lower performance of the real-valued DNN in fitting the channel mapping function, the resulting curve is relatively flat and even degraded when the number is becoming too large. Meanwhile, the Duman algorithm is merely impacted by the number of antennas at the transmitter.

Fig. 4 illustrates the relationship between the NMSE performance and the number of radiation modes per antenna when the SNR is fixed at 30 dB and the number of antennas at the transmitter are fixed at 64. 
The performance of the Duman algorithm degrades as the number of radiation modes increases. In contrast, the PR-Net outperforms the DNN, with both models producing approximately straight lines with slight fluctuations. This phenomenon can be attributed to the fact that the NMSE for each radiation mode is not exactly equal, leading to fluctuations in the average error across all radiation modes.

\section{Conclusion}
We introduce a novel channel estimation and extrapolation method for PR-MIMO systems that require the same pilot overheads compared with conventional MIMO systems. This method involves dividing the transmitter antennas into multiple groups, with each group employing different radiation mode for channel estimation. Then, we explore the channel mapping function for PR-MIMO systems and employ the complex-valued PR-Net to extrapolate channel coefficients of all radiation modes. Simulation results show that the PR-Net outperforms the existing DNN in terms of extrapolation accuracy and complexity, and deep learning methods offer significant NMSE gains over non-learning algorithm in PR-MIMO systems.

\bibliographystyle{IEEEtran}
\bibliography{citation}

\end{document}